\documentclass[aps,prl,superscriptaddress,showpacs,reprint]{revtex4-1}

\pdfoutput=1
\usepackage{color}
\usepackage{natbib}
\usepackage{amsmath}
\usepackage{amssymb}
\usepackage[pdftex]{graphicx}
\usepackage{epstopdf}
\usepackage[latin1]{inputenc}
\usepackage[ngerman, english]{babel} 
\usepackage{siunitx}
\usepackage[version =3]{mhchem}
\usepackage{soul}	
\usepackage{color}

\newcommand{\EF}{$E_{\rm F}$}
\newcommand{\EB}{$E$-$E_{\rm F}$}
\newcommand{\TN}{$T_{\rm N}$}

\newcommand{\TC}{$T_{\rm coh}$}
\newcommand{\SB}{SmB$_{\rm 6}$}

\begin{document}
\title{Universal properties of the  near-gap spectra of \SB : dynamical mean-field calculations and photoemission experiments}
\date{\today}

\author{Chul-Hee Min}
\affiliation{Experimentelle Physik VII, Universit\"at W\"urzburg,  97074 W\"urzburg, Germany}

\author{K.-S. Chen}
\affiliation{Institut f\"ur Theoretische Physik und Astrophysik, Universit\"at W\"urzburg, 97074 W\"urzburg, Germany}
\author{P. Lutz}
\affiliation{Experimentelle Physik VII, Universit\"at W\"urzburg, 97074 W\"urzburg, Germany}
\author{H. Bentmann}
\affiliation{Experimentelle Physik VII, Universit\"at W\"urzburg, 97074 W\"urzburg, Germany}

\author{B. Y. Kang}
\affiliation{School of Materials Science and Engineering, Gwangju Institute of Science and Technology (GIST), Gwangju 61005, Korea.}

\author{B. K. Cho}
\affiliation{School of Materials Science and Engineering, Gwangju Institute of Science and Technology (GIST), Gwangju 61005, Korea.}

\author{J. Werner}
\affiliation{Institut f\"ur Theoretische Physik und Astrophysik, Universit\"at W\"urzburg, 97074 W\"urzburg, Germany}
\author{F. Goth}
\affiliation{Institut f\"ur Theoretische Physik und Astrophysik, Universit\"at W\"urzburg, 97074 W\"urzburg, Germany}
\author{F. Assaad}
\affiliation{Institut f\"ur Theoretische Physik und Astrophysik, Universit\"at W\"urzburg, 97074 W\"urzburg, Germany}

\author{F. Reinert}
\affiliation{Experimentelle Physik VII, Universit\"at W\"urzburg, 97074 W\"urzburg, Germany}

\begin{abstract}
Samarium hexaboride (\SB) has been presumed to show a topological Kondo insulating state consisting of fully occupied quasiparticle bands in the concept of a Fermi liquid. This gap emerging below a small coherence temperature is the ultimate sign of coherence for a many-body system, which in addition induces a non-trivial topology. Here, we demonstrate that just one energy scale governs the gap formation in \SB, which supports the Fermi liquid description. The temperature dependence of the gap formation in the mixed valence regime is captured within the dynamical mean field (DMFT) approximation to the periodic Anderson model (PAM). The scaling property of the model with the topological coherence temperature provides a strong connection to the photoemission spectra of \SB. Our results suggest a simple way to compare a model study and an experiment result for heavy fermion insulators.

\end{abstract} 
\maketitle 

The interplay of topology and correlation effects has led to generalizations of the concept of topological insulators \cite{konig_quantum_2007,Kane_topological_2005}, that is, to symmetry protected topological states (SPT) \cite{Senthil15}. For given symmetries, it is necessary that the bulk has a unique gaped ground state but that the surface reveals gap-less edge states protected by the symmetries. Topological Kondo insulators (TKI) \cite{dzero_topological_2010,hohenadler_correlation_2013} are an example of a time reversal SPT state. Here, correlation effects are dominant in the formation of the low-energy quasiparticle excitations, but the ground state itself is believed to be adiabatically connected to a Fermi liquid. In this sense, one can make a link to non-interacting systems, and take over our understanding of the $\mathbb{Z}_2$ classification \cite{fu_topological_2007} to this class of correlated materials.

The basic understanding of the renormalized $f$\,states is substantially established in the local moment regime of the single impurity Anderson model (SIAM). When the local moments of localized $f$\,states are antiferromagnetically coupled with conduction electron states, they become quasiparticle states in the Landau Fermi liquid scheme. This local Fermi liquid phase emerges as a crossover, which smoothly happens without any phase transition \cite{haldane_scaling_1978,werner_interaction_2013,kang_topological_2015}. Moreover, the physical properties follow universal functions on one energy scale, which is the Kondo temperature $T_K$. It implies that the physical properties can be written in terms of $T/T_K$ \cite{hewson__1993}. Other crossover parameters for the paramagnetic ground states in the periodic Anderson model (PAM) have also been studied when the screened local moments, which are positioned on a lattice, form coherent Bloch-like states \cite{haldane_scaling_1978,tahvildar_low_1998,eder_many_1998,Burdin_coherence_2000,Assaad_coherence_2004,yang_scaling_2008,werner_dynamically_2014}. The coherent states can be clearly identified below the coherence temperature \TC, which is typically smaller than $T_K$ \cite{Burdin_coherence_2000, Assaad_coherence_2004}.

Because PAM may result in different ground states and phases, there are different characteristic temperatures. However, for a simple Fermi liquid ground-state, the concept of \TC~describes the emergence of the coherent quasiparticle band, especially for the \textit{mixed valence} regime \cite{haldane_scaling_1978,werner_interaction_2013,werner_dynamically_2014}. In this regime, the local moments are not clearly defined and charge fluctuations play an important role as much as spin fluctuations. Due to the charge fluctuations, spectral weight is redistributed across the gap, which has been experimentally and theoretically observed for samarium hexaboride (\SB) \cite{jdenlinger_smb6_2014,min_importance_2014,werner_dynamically_2014}. In model studies using dynamic mean-field theory (DMFT) approaches, a finite electron-electron coulomb interaction $U$, which is comparable to the single particle energy $\varepsilon_f$ of the occupied $f$\,state, is required to describe the electronic structure of the homogeneous mixed valent \SB.

Whereas qualitative and conceptual comparisons between model studies and experiments have been already made \cite{werner_interaction_2013,werner_dynamically_2014,jdenlinger_smb6_2014}, a detailed quantitative analysis is hampered by the complexity of the real system. Firstly it is difficult to extract the necessary model parameters from the complicated photoemission final state features \cite{chazalviel_study_1976,denlinger_advances_2000,jdenlinger_smb6_2014,min_two_2015}; secondly, given a minimal set of parameters, there is a considerable uncertainty in the determination of reliable and unambiguous values for each of the parameters in this large parameter space. Due to the physical nature of the crossover transition, there is, thirdly, a certain freedom in defining the actual quantitative crossover temperature, e.g. $T_K$, \TC, etc., depending on the respective physical properties \cite{haldane_scaling_1978,hewson__1993,kang_topological_2015}. As a result, it is in general rather ambiguous to connect the temperature-dependent theoretical with the experimental results for such systems. 

For the insulating cases, \TC~marks the temperature scale at which the gap opens. Thus, many physical properties will also show drastic changes at \TC, and it can be connected to the topological phase because the $\mathbb{Z}_{2}$ topological indices \cite{fu_topological_2007} are defined after the coherent renormalized $f$\,band with odd parity has formed \cite{legner_topolgical_2014,werner_dynamically_2014,chen_optical_2014}. 

Since the emergent topological band structure is nothing else but evidence for coherence, we can use the net polarization of the topological surface states to define \TC~in the model calculations \cite{yoshida_correlation_2012}. One can define a pseudo-spin Hall conductivity that takes the value of unity at zero temperature in the insulating state, thereby defining a topological index $N_2$ \cite{werner_dynamically_2014,chen_optical_2014}. Topological coherence temperature \TN~corresponds to the temperature scale at which the index $N_2$ takes the value of half, \textit{i.e.} the pseudo-spin of the topological surface states becomes half of the full net polarization. With the definition of \TN, a unifying crossover parameter can be defined to describe both the lattice coherence and the topological phase. Moreover, scaling behavior with \TN~can appear for both the local-moment and the mixed-valence regimes \cite{werner_dynamically_2014}.

Thus, we would like to investigate the existence of the energy scale in the photoemission spectra of \SB. \SB~is a paramagnetic bulk insulator with a smooth gap opening related to $f$\,bands \cite{allen_large_1979,wolgast_low-temperature_2013}. Consistency might be obtained by finding an appropriate energy scale shared by both theory and experiment. After showing the results of the scaling, we will discuss the indication of the scaling with \TC.

Theoretical studies on the temperature dependence of the electronic structure for a mixed valent insulator in 2D square lattice were carried out based on the DMFT. The numerically exact CT-HYB quantum Monte Carlo algorithm is applied to solve the auxiliary impurity problem. For the study of the mixed valence regime, we use tight binding bands with the same input parameters as in the Ref\,\cite{werner_dynamically_2014}, which are normalized with the conduction hopping parameter $t$, e.g. $U/t=5.0$, $\varepsilon_f/t=-6.0$, the hybridization $V/t=0.4$, etc. The parameter set gives \TN$/t=0.21$. The high resolution photoelectron spectroscopy (PES) experiments were carried out at the UE112-PGM-1b (``$1^{3}$'') beamline of BESSY~II using a Scienta R4000 analyzer at 3\,K\,$\leq$\,$T^{exp}$\,$\leq$\,60\,K \cite{min_importance_2014}. The excitation photon energy was $h\nu$\,$=$\,70\,eV, whose constant energy map at \EF~covers $k_{z}$\,=\,6$\pi/a$ in normal emission, with the energy resolution of 7\,meV. 

\begin{figure}
	\centering
		
\includegraphics[width=0.5\textwidth]{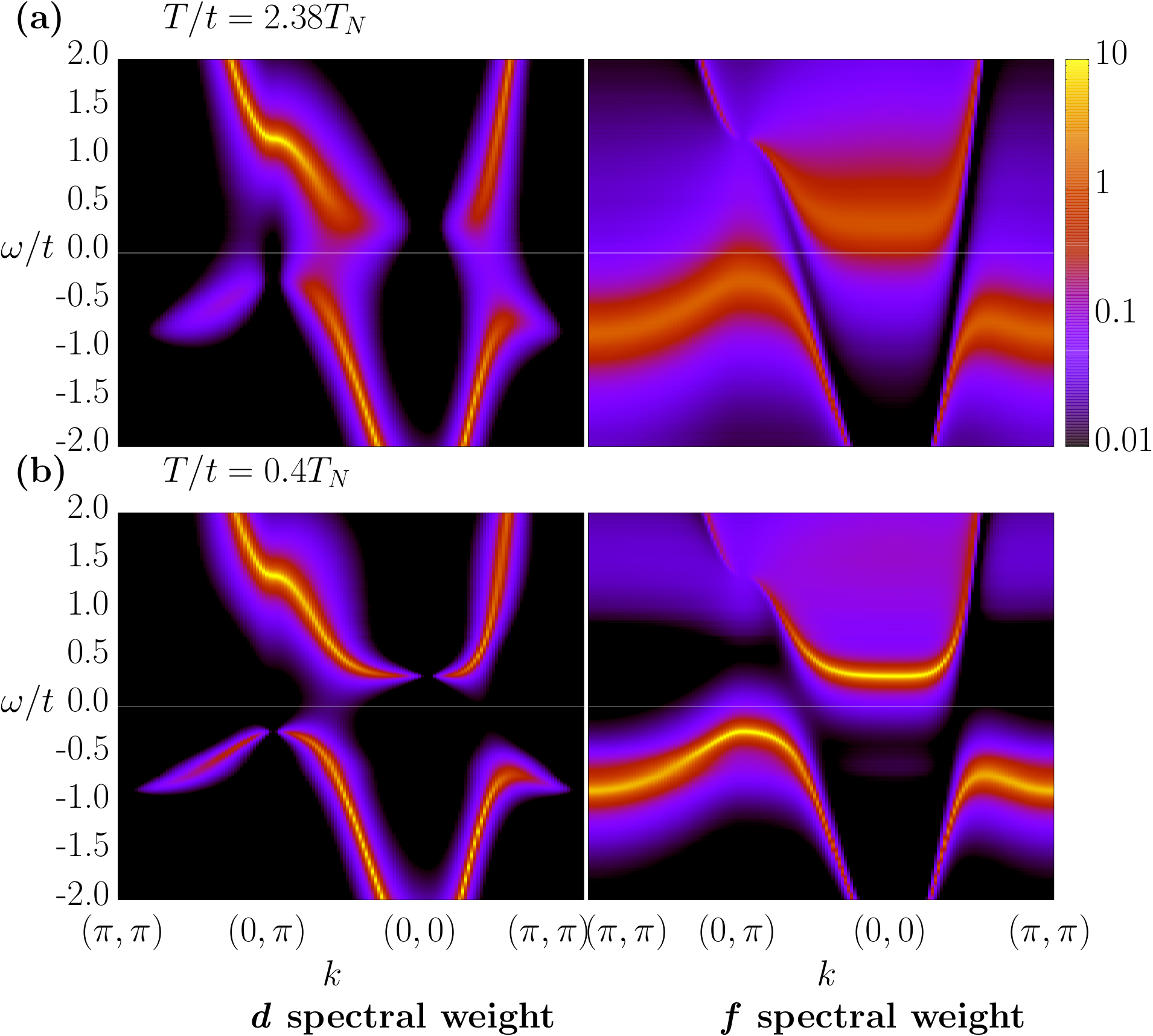}

			\caption{(Color online). Single-particle spectral functions in the mixed valence regime above and below the topological coherence temperature \TN. Left and right panels show the conduction $d$\, and localized $f$\,spectral weight, respectively. (a) Above \TN, the $f$\,spectral weights are strongly diffused. The blurred $f$\,features strongly contribute to the total spectral weight in the gap region. (b) Below \TN, the $f$\,spectral weight in the gap region is dramatically reduced, and the $f$\,dispersion becomes clearer. Moreover, the dispersions of the $d$ and $f$\,features become identical, which is a result of the strong character mixing.}
	\label{fig:fig1}
\end{figure}
Fig.\,\ref{fig:fig1} shows how signatures of coherency appear in the mixed valent insulator by comparing the theoretical spectral functions at temperatures above and below \TN. The temperature dependent $d$ and $f$\,spectral weights obtained from DMFT studies are shown on the left and right panels, respectively (See Ref.\,\cite{werner_dynamically_2014,chen_optical_2014} for details). Two main changes appear as one crosses the temperature scale  \TN, Fig.\,\ref{fig:fig1}\,(a) and (b). One is that the gap is getting more pronounced, and the other is that the $f$ and $d$\,spectra becomes more similar to each other.
Close to \EF, the $f$\,features are getting blurred at $T$\,$>$\,\TN~(Fig.\,\ref{fig:fig1}\,(a, right)), whereas they have reduced intensity at $T$\,$<$\,\TN~(Fig.\,\ref{fig:fig1}\,(b, right)). The reduced spectral weight in the gap region has been already observed in other PES investigations \cite{souma_direct_2002, nozawa_ultrahigh-resolution_2002,jiang_observation_2013,xu_exotic_2014}. In addition, the band dispersions of heavy $f$\,states and itinerant $d$\,states change as a function of the temperature. Above \TN, $d$ and $f$\,spectral weights are strongly dispersing and flat bands, respectively, whereas below \TN~they become more similar to each other. This is due to hybridization, which gives rise to the character mixing of $f$ and $d$. We note that in the model calculations the hybridization between the $f$ and $d$  states is given by $\pmb{V}(\pmb{k}) \cdot \pmb{\sigma}$, where $\pmb{\sigma}$ is a vector of Pauli spin matrices. Time reversal symmetry requires $\pmb{V}(\pmb{k})$ to be odd, such that the hybridization vanishes at time reversal momenta. This explains in particular the loss of spectral weight at the high symmetry points $(0,0)$, $(0,\pi)$ and $(\pi,\pi)$ at low temperatures.

\begin{figure}
	\centering
		\includegraphics[width=0.5\textwidth]{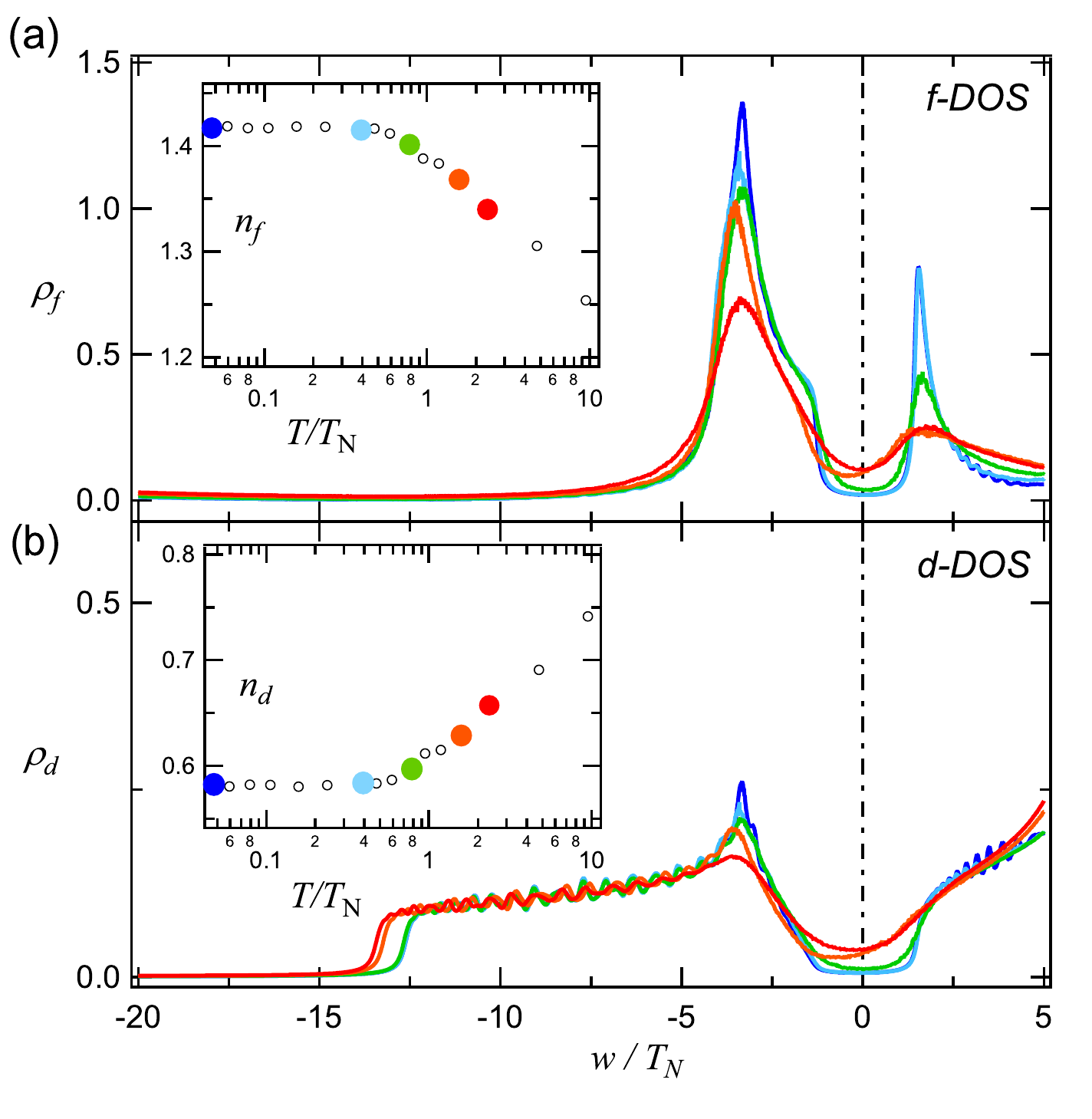}
			\caption{(Color online). Temperature dependence for density of $f$ and $d$\,states are shown in Fig.\,\ref{fig:fig2}\,(a) and (b), respectively. The energy is scaled with \TN. In the same colors used for DOS lines, the corresponding occupation numbers for $f$ and $d$\,states are shown as a function of $T/$\TN~in the insets. (a) As seen already in Fig.\,\ref{fig:fig1}, the gap, whose character mainly originates from $f$\,states at high temperatures, becomes clearer and deeper with decreasing temperature. Note that at $T$\,$\leq$\,\TN~the spectral weights inside the gap, and the occupation numbers of $f$ and $d$\,states start to saturate (insets).}
		\label{fig:fig2}
\end{figure}
Fig.\,\ref{fig:fig2} shows the temperature-dependence of the $f$ and $d$ density of states (DOS) as a function of energy scaled by \TN~($w/$\TN) \cite{werner_dynamically_2014}. In the insets, the occupation numbers $n_f$ and $n_d$ of each DOS are depicted as a function of $T/$\TN~in the colors corresponding DOS lines. Fig.\,\ref{fig:fig2}\,(a) shows the occupied $f$\,-DOS near \EF, which is related to the upper Hubbard band, and Fig.\,\ref{fig:fig2}\,(b) shows the occupied $d$\,-DOS, which originates from the 2D conduction band. Both $f$- and $d$\,-\,DOS are only broadened due to the imaginary part of the self-energy. (The wiggling features in Fig.\,\ref{fig:fig2}\,(b) are reminiscent of finite k-mesh used in the DMFT calculations). The spectral weight in the gap region ($\lvert{w/T_{N}}\rvert\,\leq\,1$) clearly reduces with decreasing temperature. We note that at $T$\,$=$\,\TN~the intensity at \EF~is saturated in both $f$- and $d$-\,DOS. Moreover, the occupation numbers of $f$ and $d$\,states are almost saturated to the maximum and minimum occupations, respectively. Thus, \TN~is the crossover point when the gap is clearly open while the charge transfer between $f$ and $d$\,states is saturated because both states are strongly intermixed.

Based on the results from theory, we search for a corresponding energy scale of ${T}_{\rm coh}^{exp}$~for \SB~ as determined from experiments. In addition to recent high resolution PES studies showing the gap opening below 60\,K \cite{min_importance_2014,denlinger_temperature_2013}, various other experiments have revealed a similar characteristic temperature of $\sim$\,50\,K \cite{menth_magnetic_1969,Sluchanko_nature_1999,  sluchanko_intragap_2000, sluchanko_low_2001,Zhang_hybri_2013, Roessler_hybridization_2014,cooley_smb_6:_1995, wolgast_low-temperature_2013,kasuya_valence_1979,Phelan_Correlation_2014,Ruan_emergence_2014,pena_nmr_1981,Takigawa_nmr_1981,mizumaki_temperature_2009,Nyhus_low_1997,Roman_transport_1997,Caldwell_high_2007,Yeo_effects_2012,Biswas_low_2014,allen_large_1979}. In particular, the signature of ${T}_{\rm coh}^{exp}$~can be estimated from the XAS results by taking the saturation point ($\sim$\,50\,K) in the temperature-dependent Sm valence \cite{mizumaki_temperature_2009}. Since \TN~is positioned near the saturation region as shown in the insets of Fig.\,\ref{fig:fig2}, the coherence temperature of \SB~is determined to be ${T}_{\rm coh}^{exp}$\,$\sim$\,50\,K.

\begin{figure}
	\centering
				\includegraphics[width=0.5\textwidth]{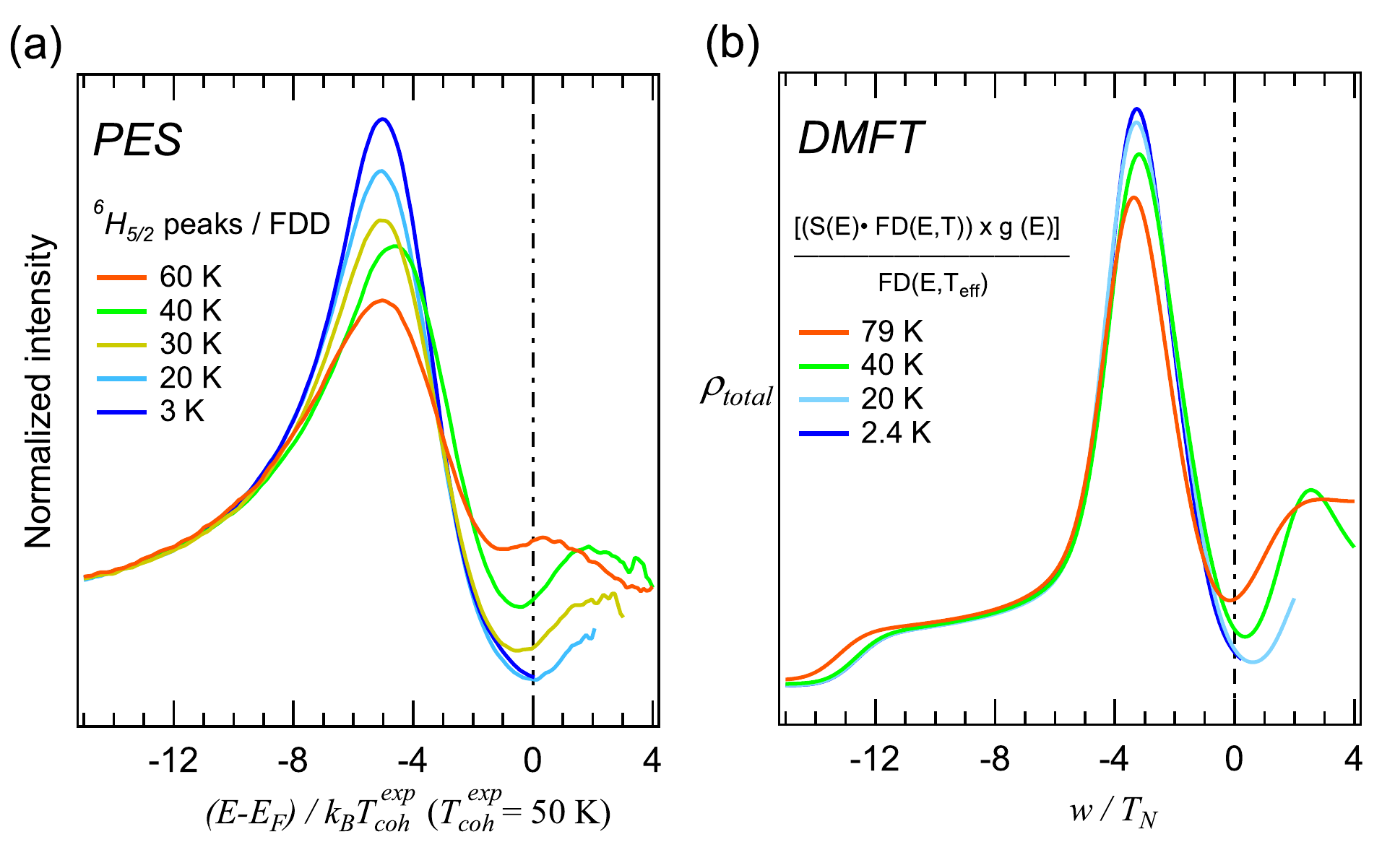}
				\caption{(Color online). Comparison of the PES spectra of \SB~with the theoretical DOS after  scaling with the corresponding \TC. (a) The angle-integrated spectra of \SB~are divided by the Fermi-Dirac function \cite{min_importance_2014}. The energy axis is reduced by the coherence scale determined from experiments (${T}_{\rm coh}^{exp}=50 K$). (b) Reconstructed spectra from theoretical calculation, considering Fermi-Dirac distributions and total experimental resolutions (see text), in order to compare with the experiment spectra (a). The $f$\,peaks in (a) and (b) appear at a similar position. Moreover, the gap regions show similar temperature dependence.}
				\label{fig:fig3}
\end{figure}
Fig.\,\ref{fig:fig3}\,(a) shows the angle-integrated photoemission spectra of \SB~divided by the Fermi-Dirac distribution (FDD) \cite{min_importance_2014} on the reduced energy scale of ${T}_{\rm coh}^{exp}$. The peak at (\EB)$/(k_B{T}_{\rm coh}^{exp})=-5$ is the $4f$\,multiplet excitation ($^{6}H_{5/2}$) with lowest binding energy. The gap in the PES data is getting deeper with decreasing temperature, comparable to the behavior observed in the theoretical DOS in Fig.\,\ref{fig:fig2}. We compare now the experimental and the theoretical spectra in more detail using their functional relation to the scaling. If \SB~has the scaling property, the energy axis of the DOS can be rescaled by setting $w$/\TN\,$=$\,(\EB)/($k_B$${T}_{\rm coh}^{exp}$) to link with the experimental energy. Moreover, theoretical temperatures can be matched with experimental temperatures $T^{exp}$ by writing $T$/\TN\,$=$\,$T^{exp}$/${T}_{\rm coh}^{exp}$.

To make the calculated DOS directly comparable to the experimental spectrum, we treated the DOS by a numerical procedure already successfully applied to metallic systems \cite{ehm_quantitative_2002}. The total DOS, which is the sum of $f$- and $d$-\,DOS, was multiplied by the FDD, and convoluted by a Gaussian function to consider the experimental broadening due to finite energy resolution. Finally, the resulting spectrum was normalized to the FDD as shown in Fig.\,\ref{fig:fig3}\,(b) in order to get comparable spectra to the ones in Fig.\,\ref{fig:fig3}\,(a).

As a result, the scaled theoretical and experiment spectra show a surprisingly good agreement in the temperature dependence of both the $4f$\,peak and the gap opening. Note that the energy positions and the line widths of the $4f$\,peaks, and the size of  the gap are on the scale of the respective \TC, which was already theoretically suggested \cite{werner_dynamically_2014,chen_optical_2014}. Moreover, above 60\,K both gap centers appear to be below \EF~and shift closer to \EF~with decreasing temperature, and below 20\,K both spectral weights at \EF~saturate  for both theory and experiment.

Thus, with the theoretical and experimental \TC, \textit{i.e.} \TN~and ${T}_{\rm coh}^{exp}$, the scaling yields the quantitative connection between the model calculation and the photoemission spectra. In light of the very crude choice of model parameters, this result is very impressive because it shows that few key parameters, namely the \textit{mixed valence configuration}, and the \textit{coherence temperature}, suffice to provide an account of the experiments. It indicates the existence of one energy scale during the gap formation, which involves the emergence of the coherent renormalized bands that is a Fermi liquid. This results in the universal temperature dependence of a mixed valent insulator \SB. 

We have searched for correspondence between the DMFT and PES spectra of \SB~to understand the correlation-driven emergence of the coherent $4f$\,features and the concomitant gap formation. The comparison relies on the following four points: First, we have used the photon energy $h\nu$\,$>$\,40\,eV to emphasize the $4f$\,character in the photoemission spectra, which takes into account the dominant spectral weight of $f$\,states near gap region in the model study. Second, we have considered only the state $^{6}H_{5/2}$ appearing at the lowest energy below \EF. Experimentally, the next lowest state $^{6}H_{7/2}$, which appears at the energy of \EB\,$\approx-0.18$\,eV \cite{chazalviel_study_1976,denlinger_advances_2000,jdenlinger_smb6_2014,min_two_2015}, shows much less changes in both shape and intensity than $^{6}H_{5/2}$, and thus shall be neglected for the contribution to low-temperature behavior. Third, as \TC, defining the onset of the gap opening, we selected a prominent characteristic temperature found also in various experiments on \SB~\cite{menth_magnetic_1969,Sluchanko_nature_1999,  sluchanko_intragap_2000, sluchanko_low_2001,Zhang_hybri_2013, Roessler_hybridization_2014,cooley_smb_6:_1995, wolgast_low-temperature_2013,kasuya_valence_1979,Phelan_Correlation_2014,Ruan_emergence_2014,pena_nmr_1981,Takigawa_nmr_1981,mizumaki_temperature_2009,Nyhus_low_1997,Roman_transport_1997,Caldwell_high_2007,Yeo_effects_2012,Biswas_low_2014,allen_large_1979}, namely 50\,K. And, fourth, to make contact with the DMFT calculation, we have chosen model parameters to obtain a system in the mixed valence regime, where the occupancies still vary at temperature near gap opening. 

In summary, we have studied the characteristic spectral features of \SB~by simple DMFT model calculations and by photoemission spectroscopy. By scaling the data with the respective coherence temperature, we have obtained a considerable quantitative agreement between theory and experiment, although the details of the band structure of \SB~have not been considered in the calculations. The agreement gives strong evidence for the existence of a scaling behavior in the electronic structure of \SB, which indicates that the formation of the coherent quasiparticle states is essential for the gap opening.

We thankfully acknowledge stimulating discussions with H.-D. Kim, and J.-D. Denlinger. This research was supported by the DFG (through SFB 1170 "ToCoTronics", projects A01, C01, C06, Z03). B.K.C. and B.Y.K. were supported by National Research Foundation of Korea (NRF) grants funded by the Korean government (MSIP; Grants No. 2011-0028736 and Bank for Quantum Electronic Materials-BQEM00001).

%

\end{document}